\documentclass[10pt,aps,superscriptaddress,twocolumn,showpacs,preprintnumbers,prb]{revtex4-1}

\usepackage{graphicx,epsfig,color}
\usepackage{times}

\begin{document}
\title{Using Kondo entanglement to induce spin correlations between disconnected quantum dots} 
\author{C. A. B\"usser}
\affiliation{Instituto de Ciencias B\'asicas y Experimentales, Universidad de Mor\'on, Buenos Aires, Argentina  }
\email{Corresponding author: busserc@gmail.com}
\begin{abstract}
We investigate the entanglement between the spins of two quantum dots that are not connected at once to the same system. Quantum entanglement between localized spins is an essential property for the development of quantum computing and quantum information. It is for this reason that generating and controlling an entangled state between quantum dots received great attention in the later years. 

In this work, we show that the information on the spin orientation of a quantum dot can be kept, using the Kondo entanglement, in a reservoir of electrons. Then, this information can be transmitted to another dot after the first dot has been uncoupled from the reservoirs. We use a double quantum dot system in a parallel geometry to construct the initial state, where each dot interacts with reservoirs of different symmetries. We chose a phase in the couplings to induce an antiferromagnetic spin correlation between the dots. The time evolution of the initial state has been analyzed using the time-dependent density matrix renormalization group method.

We found that a partially entangled state between the dots can be obtained, even if they are not connected at the same time. This entangled state is formed just during the transient and is destroyed in the stationary state. The stability of the state found in the transient is shown. To understand the details of these phenomena, a canonical transformation of the real space is used.
\end{abstract}
\maketitle

\section{Introduction:}

One of the most surprising effects of quantum mechanics is the entanglement between two systems.\cite{horodecki09,*Chen18,*Zou21}
This effect describes a nonlocal correlation between different quantum objects.
The possibility of controlling an entangled state is of great importance to the development of quantum computation and quantum information and has received much attention in recent years.\cite{Nielsen10}
The physical manipulation of entangled electrons in systems composed of quantum dots (QDs) is a central problem for the possibility of developing a solid-state qubit.\cite{Reilly15,*Lavroff21}
On the other hand, the study of the electronic transport through systems of several QDs is fundamental for the development of the industry of single-electron devices.\cite{sasaki00,vanderwiel02,*Zwanenburg13,*Akinwande19,Guo21}

An interesting question that can be raised is whether the information about the spin projection of an electron located at a quantum dot can be stored in a metal reservoir. The objective of this effort is to show that, at least for a short time, this can be done.
To demonstrate that this outcome can be done, we use the Kondo effect\cite{hewsonbook,*goldhabergordon98,*kouwenhoven01}, where a singlet state between a localized spin and a free electron reservoir is formed. 
Within the metallic lead emerges a structure that shields the localized spin of the remaining spins of the electronic sea. This structure is generally known as {\it Kondo cloud}.\cite{busser10,*ghosh14,*laercio19,*borzenets20}
In this way, this effect may be regarded as an entanglement between a local spin and a sea of delocalized spins.\cite{feiguin17,bonazzola17,yoo18}
Once the interaction that produces the Kondo effect is removed, and once a certain relaxation time has passed, the Kondo cloud structure disappears into the electron sea and the Kondo singlet is destroyed. 
In this article, we show that the spin information stored in the Kondo cloud can be used to transfer the information on the spin orientation of the localized spin to another system.

The rise of quantum computation and quantum information has increased the interest in the study of coupled quantum dots, mainly because the spin in a QD can act as a qubit~\cite{Maslova17}. In particular, the study of two QDs connected to metal leads due to the possibility to create entangled states. These special states are mostly analyzed in the stationary case, but many of their properties can raise as time-dependent non-stationary characteristics.~\cite{Maslova18} Just to cite some of these characteristics, we can mention that an entangled state between QDs can be generated through a Ruderman-Kittel-Kasuya-Yoshida interaction (RKKY)~\cite{kittel,*Kasuya,*Zhang19} just applying a bias potential~\cite{busser13a,Hou15} and also can be generated using bound states in the continuum and vibrational states~\cite{Alvarez15,Maslova20}.
Systems with 3QDs had also attracted attention due to the possibility to have, as a function of the number of electrons, a magnetic frustration rather than an entangled state.\cite{Vernek09}
Among the numerical methods used to calculate the electronic properties of quantum dots, we can mention the equation of motion (EOM)\cite{Lacroix81}, numerical renormalization group\cite{NRG}, density matrix renormalization group\cite{Feiguinbook}, functional renormalization group\cite{Andergassen10}, embedded-cluster approximation\cite{anda08}, just to name a few. Recently, an extension of EOM called hierarchical equation of motion (HEOM) had been introduced with success to calculate the electronic properties as well as the time evolution of QDs connected to metal reservoirs.\cite{Hou15,Gong18,Zhang23}

The system used to study this effect consists of two quantum dots connected to two metal leads with a parallel Aharanov-Bohm geometry. Choosing the correct phase in the interferometer, the spin correlation between the electrons located at the dots is zero. This is the initial state prepared in the two settings discussed on these pages.
In the first setting, we review the results of the spin correlation between the QDs induced by a bias potential.\cite{busser13a,Hou15}
For the second setting considered, we show that the information on the spin projection of an uncoupled dot can be transferred to the QD that remains coupled. This can be done, during a short period, taking advantage of the Kondo singlet formed with the metallic reservoirs. 
To understand the results of these two settings, we use an even-odd canonical transformation of the degrees of freedom in the metal leads\cite{jones87,feiguin12a}. This class of transformation had been proposed to study the phases in spin chains and problems where an impurity is connected to a ring.\cite{busser12b,Barcza19,buruaga19}

This work is organized as follows. In the next section, we present the model Hamiltonian for the two quantum dots. In Sec.~\ref{previous} we review the results for two dots in an Aharanov-Bohm interferometer. Next, in Sec.~\ref{results}, we present the case when one of the quantum dots is decoupled before applying a bias that connects the  different symmetries of the reservoirs. In Sec.~\ref{quench} we show the stability of the entangled state, and we study the decoherence process. Finally, in Sec.\ref{conclusion}, we present our conclusions.

\section{ Model for two Quantum Dots}\label{model}

\subsection{Hamiltonian Model}

\begin{figure}
\centerline{\epsfxsize=8.0cm\epsfbox{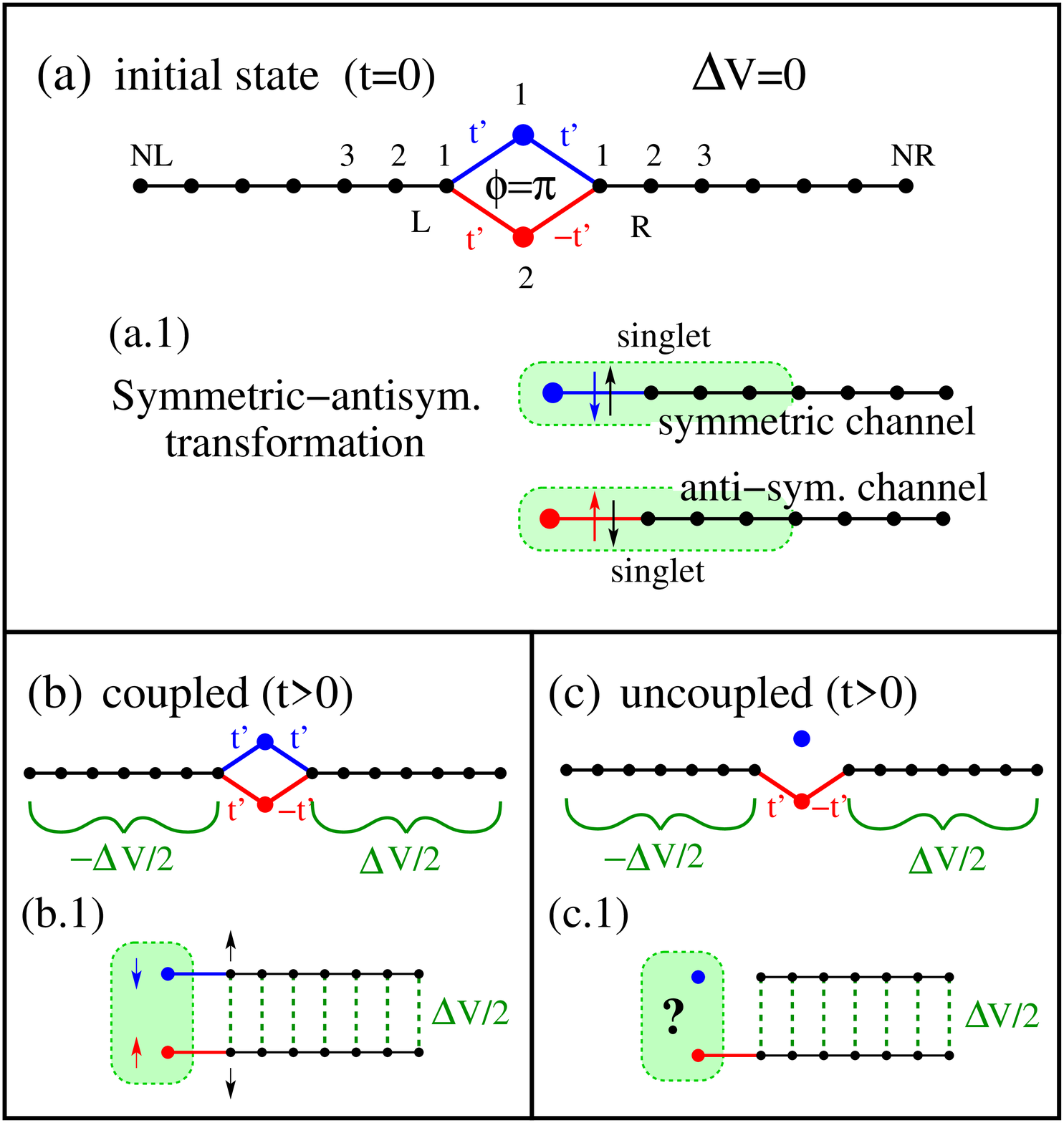}}
\caption{(Color online) Schematic of  the two quantum dots (QDs) system and the canonical transformation Eqs.~(\ref{eq:trafo1}) and~(\ref{eq:trafo2}).  
(a) the initial state for the DMRG calculations, (b) the Aharanov-Bohm interferometer structure and (c) the case of the uncoupled dot analyzed in sections~\ref{results} and~\ref{quench}.
To understand the results found by tDMRG a canonical transformation, that decoupled the problem in two parts (symmetric and antisymmetric), is applied. The results of this transformation are presented for each case in panels (a.1), (b.1) and (c.1).
}
\label{figure0}
\end{figure}

We use the Anderson impurity model (AIM) to represent two parallel quantum dots (QDs) connected to different leads in an Aharanov-Bohm geometry. The resulting Hamiltonian is,
\begin{eqnarray}
H &=& H_{\rm leads} + H_{\rm hy} + H_{\rm int}  , \label{Htotal} \\
H_{\rm leads} &=& \sum_{\alpha=L,R}~~\sum_{i=1;\sigma}^{N-1} \left[ -t_0 (c_{\alpha i\sigma}^\dagger c_{\alpha i+1\sigma}+h.c.)\right] \nonumber \\
	&& +  \sum_{i=1;\sigma}^N  \left[ \mu_L n_{L i \sigma}   +  \mu_R n_{R i \sigma} \right] ,  \label{leads} \\
H_{\rm int} &=& \sum_{j=1,2 ;\sigma} [U n_{j\sigma}n_{j\bar{\sigma}}  + V_g n_{j\sigma} ] \\     		
H_{\rm hy} &=& -t'_1 [d^\dagger_{1\sigma} c_{L1\sigma} + c_{R1\sigma}^\dagger d_{1\sigma} +h.c.]~+~ \nonumber \\
&& -t_2' [d^\dagger_{2\sigma} c_{L1\sigma} +  {\displaystyle\mbox{e}^{i\varphi}} c_{R1\sigma}^\dagger d_{2\sigma}+h.c.] \label{Hphase} . 
\end{eqnarray}
This Hamiltonian is made up of three parts. The first part, $H_{\rm leads}$, which describes the reservoirs, is composed of two chains of length $N$ of non-interacting sites. These two chains (labeled $R$ and $L$), characterize the non-interacting right and left metallic leads. The constant hopping matrix element $t_0=1$ inside the chains is used as the unit of energy. In the middle, between these chains, are two quantum dots labeled by $j=1,2$. 
The second term of the Hamiltonian, $H_{\rm int}$, describes the many-body interactions inside each dot as well as the gate potential. The $U$ term is the Coulomb repulsion when are two electrons inside any QD; the $V_g$ term is a gate potential chosen to have the QDs at half filling.
The last term $H_{\rm hy}$ is the hybridization between the localized levels of the dots and the leads.
We define the tunneling strength by $\Gamma= 2\pi t'^2 \rho_{\rm leads}(E_F)= 2t'^2$, where $\rho_{\rm leads}(E_F)$ is the local density of states (LDOS) of the leads. If the leads are described by a semi-infinite chain of sites at half filling, we have that $\rho_{\rm leads}(E_F)= 1/\pi$. Then $\Gamma=2t'^2$ for the thermodynamic limit.

We incorporate an arbitrary fixed phase $\varphi=\pi$ in the hopping matrix element between the second dot and the right lead. The origin of this phase can be associated with a magnetic flux in an Aharanov-Bohm interferometer or two degenerated levels of a single QD coupled with different symmetries to the reservoirs.\cite{busser13a,Hou15}

In this article, we consider fully symmetric tunnel couplings, i.e.,  $|t'_1|=|t'_2|=t'$. See the supplemental material in Reference~[\onlinecite{busser13a}] for the discussion when $|t'_1|\neq|t'_2|$.

Observe that in Eq.~(\ref{leads}) are included two terms  $\mu_L$ and $\mu_R$ that mimic the chemical potentials of the leads when a bias is applied. We apply the bias $\Delta V$ symmetrically between the leads, thus $\mu_L=\Delta V/2$ and $\mu_R=-\Delta V/2$. This bias $\Delta V$ induces a current from left to right leads.

In Eq.~\ref{Htotal} we use the common notation for operators. The operator $c_{\alpha l\sigma}^\dagger$ ($c_{\alpha l\sigma}$) creates (annihilates) an electron with spin $\sigma$ at site $l$ in the $\alpha=L,R$ lead; operator $d_{j \sigma}^\dagger$ ($d_{j\sigma}$) creates (annihilates) an electron on QD$j$. Finally, the operator $n_{\alpha l \sigma}= c_{\alpha l \sigma}^\dagger c_{\alpha l \sigma}$ is the number of particles. 

The total size of the system is $2N+2$.
In Figure ~\ref{figure0}(a) we present a schematic representation of the proposed system.

\subsection{Spin correlations and Current}

Due to the 1D symmetry of the Hamiltonian proposed in the previous section, we chose the Density Matrix Renormalization Group method (DMRG)\cite{vidal04,*white04,Feiguinbook} to study this problem.
The ground state  and the linear conductance of  systems  described by the AIM were extensively studied in Ref.~[\onlinecite{hofstetter01,*apel04,*zitko06,*zitko06b,*lopez07,*zitko12,zitko07}]. Related QDs model with a finite flux $\varphi$ and with spin-polarized electrons was discussed in Ref.~[\onlinecite{boese01,*koenig02,*meden06,*kashcheyevs07,*bedkihal12}].

In this work we proceed as follows, first, we use DMRG to obtain the ground state $|\Psi_0\rangle$. Then, we calculate the effect of a finite bias voltage $\Delta V$ by time-evolving the wave function $|\Psi(t)\rangle$ and measuring its properties such as the current and spin correlations as a function of time $t$. This method has been successfully used to study non-equilibrium transport through nanostructures with many-body correlations \cite{alhassanieh06,*kirino08,*boulat08,*branschaedel10a,*einhellinger12,*nuss13}. 
We evaluate the spin correlations from,
\begin{equation}
\mbox{S}_{12}(t) = \langle \Psi(t) | \vec{S}_1 \cdot \vec{S}_2  |\Psi(t) \rangle \,.
\end{equation}
The current between two sites in the leads is defined by\cite{alhassanieh06,alhassanieh09}
\begin{equation}
J_{l,m}(t) =  i t_{0} \sum_\sigma \langle \Psi(t) |c^\dagger_{l\sigma}c_{m\sigma} - c^\dagger_{m\sigma}c_{l\sigma} | \Psi(t) \rangle \,.  
\end{equation}

We work with the current averaged over the first link in the left and right lead as, 
\[ J=\frac{J_{L2,L1}+J_{R1,R2}}{2} \]

Our calculations start, at the time $t=0$, from the system in equilibrium with finite couplings $t'_1=t'_2$ and a charge per dot of $\Delta V=\mu_L-\mu_R$ that drives the system out of equilibrium. The two quantum dots are treated as a  super-site, permitting the use of a Trotter-Suzuki breakup of $\exp{(-iHt)}$.\cite{schollwoeck05,*schollwoeck11} 
The time step is $\delta t\sim 0.1$, and we enforce a fixed discarded weight of $10^{-5}$ or less, keeping a maximum of 4000 to 6000 DMRG states. All runs are performed at an overall half-filling of dots and leads.

We focus our attention on two different cases. In the first case, we keep both dots connected after applying the bias potential. To introduce the effect that is relevant in the main part of the article, we begin reviewing the results of reference~\onlinecite{busser13a,busser14}. Next, in the second case, we study what happens when we decouple one of the QD from the reservoirs before applying the bias potential. This is the central part of our study.

In the initial state of both cases, we use  $t'_2=t'_1=t'$ and $\varphi=\pi$ in Eq.~\ref{Hphase}.
As will be discussed in the next section, this phase induces an antiferromagnetic (AF) correlation between the spins of QDs.

\begin{figure}
\centerline{\epsfxsize=8.5cm\epsfbox{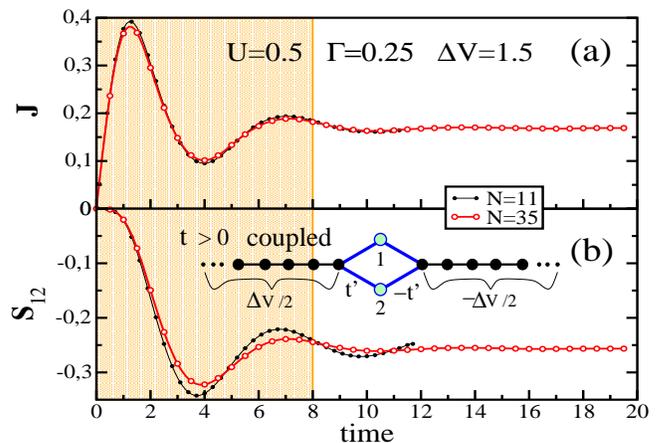}}
\caption{(Color online) Typical results for current $J(t)$ and spin correlations $S_{12}(t)$ as a function of time. In panel  (a) we present currents for the system with the 2QDs coupled; a steady current, shown by a plateau, is found after a transient. In panel (b) Spin correlation between the 2QDs; note that, after a transient, a steady antiferromagnetic correlation develops. The shaded region indicates the transient regime for the current. }
\label{figure1}
\end{figure}

\section{The Aharanov-Bohm interferometer and a canonical transformation}\label{previous}

As mentioned before, in this article we present the study of two systems; one with both dots connected all the time and the other with one of the dots uncoupled when the bias potential is applied. We refer s the {\it coupled system} to the first and {\it uncoupled system} to the second. This section is dedicated to reviewing the known results of the coupled system.\cite{busser13a,busser14,Hou15}
These calculations were done using time depending DMRG (tDMRG)~\cite{DMRGt},  master equations~\cite{deVega17} and verified using HEOM~\cite{Hou15,Gong18}

In this work, we are focused on the AF correlation for $\mbox{S}_{12}$, and for that reason, we just review the case $\varphi=\pi$. The case $\varphi=0$ presents just ferromagnetic correlations between the QDs and is discussed in Ref.~\onlinecite{busser13a}.
  
In Fig.~\ref{figure1}(a) and (b) we present a typical current and spin correlations for the two connected dots as a function of time.
Both present two regimes; first a transient state, in this case, up approximate time $\sim 8$, and then a steady state where the value of the current and the spin correlations became almost constant. 

An unexpected behavior is observed for the spin correlations $\mbox{S}_{12}$. For time $t=0$, $\mbox{S}_{12}$ is zero, and when the bias potential is applied an antiferromagnetic correlation, with value $-0.25$, emerges and became stable at the steady state regime.

The qualitative behavior of the spin correlations can be  understood using a canonical transformation of the operators of the leads, which is given by (see, e.g., Refs. \onlinecite{jones87,*ingersent92,*feiguin12a,boese01,*koenig02,*meden06,*kashcheyevs07,*bedkihal12}):
\begin{eqnarray}
c_{{\rm s} l \sigma} &=&  (c_{\rm R l \sigma} + c_{\rm L l \sigma})/\sqrt{2}, \label{eq:trafo1} \\
c_{{\rm a} l \sigma} &=&  (c_{\rm R l \sigma} - c_{\rm L l \sigma})/\sqrt{2}, \label{eq:trafo2} 
\end{eqnarray}
where $\rm s, a$ are the symmetric and antisymmetric combinations, respectively.
The result of this transformation is sketched in Fig.~\ref{figure0}~(a.1), where the leads shown represent the new states obtained from Eqs.~(\ref{eq:trafo1}) and~(\ref{eq:trafo2}).
After the canonical transformation, the Hamiltonian with $\varphi=\pi$ results in, 
\begin{eqnarray}
H_{\rm leads} &=& \sum_{\alpha={\rm s,a}}\sum_{i=1;\sigma}^{N-1}  -t_0 (c_{\alpha i\sigma}^\dagger c_{\alpha i+1\sigma}) \nonumber \\
	&& + \frac{\Delta V}{2} \sum_{i=1;\sigma}^N    c_{{\rm s} i\sigma}^\dagger c_{{\rm a} i\sigma}  ~+h.c. ,  \label{leads2} \\
H_{\rm hy} &=& -\sqrt{2}~t' ~d^\dagger_{1\sigma} c_{{\rm s} \sigma} ~-~  \sqrt{2}~t' ~d^\dagger_{2\sigma} c_{{\rm a} \sigma}~ +h.c.  
\end{eqnarray}
where we have used that $\mu_L=\Delta V/2$ and $\mu_R=-\Delta V/2$. We want to stress that, as this transformation just affects the leads operators, the QDs Hamiltonian, $H_{\rm int}$, is not modified. 

Observe that the Hamiltonian $H_{\rm leads}$ have now a ladder geometry where the bias potential $\Delta V$ appears on the rungs.

In the absence of a bias voltage, there is no direct coupling between these new states, as depicted in  Figs.~\ref{figure0}(a.1). 
Most important, QD1 is coupled to the symmetric states and QD2 to the antisymmetric states. Then, for $\varphi=\pi$, the dots are part of two decoupled subsystems and, therefore, $S_{12}$ vanishes.

Upon applying a bias, one obtains a ladder geometry where the voltage acts as a transverse coupling between the symmetric and antisymmetric states of Eqs.~(\ref{eq:trafo1}) and~(\ref{eq:trafo2}) as shown in Fig.~\ref{figure0}(b.1).

For $\varphi=\pi$ and $\Delta V\not=0$, the dots are now connected through paths with an even number of sites in the effective leads, and therefore, in the {\it ground state} of such a geometry, one expects a finite negative spin correlation. Our numerical results, shown in Fig.~\ref{figure1}(b), reveal that this same behavior appears in {\it non-equilibrium} as well.
Note that a fine-tuning of the coupling parameters is not necessary to observe an AF $S_{12}$ induced by a bias $\Delta V$.\cite{note2}

In Fig.~\ref{figure1}~(a) and (b) we can observe that the steady state is independent of the size of the leads $N$; but, on the other hand, in the transient regime, there is a strong size effect. This size dependency is due to the Kondo effect, where the sizes shown in the figure are not large enough to describe properly the Kondo ground state. In the steady state, the Ruderman-Kittel-Kasuya-Yoshida interaction mediated by the bias eliminates the Kondo effect and its size effect.\cite{busser13a}. We will come back to this issue with more detail in the next section.  

To see the connection between the spin correlation $S_{12}$ and the entanglement, we use the concurrence.\cite{hill97,*kessler13,feiguin17} 
A value of the concurrence close to one ensures a maximum entanglement, on the other hand, a concurrence with a zero value means that there is no entanglement at all. In this latter case, an AF spin correlation can be constructed using a charge fluctuation.
In the steady state of Fig.~\ref{figure1}, the concurrence is $\sim 0.3$, which shows a partially entangled state.
For a further discussion of concurrence in this system, see reference~\onlinecite{busser13a}.

To complete this review, let's mention that for $\varphi=0$, the RKKY interaction gives rise to a ferromagnetic correlation between the dots since each path that connects them involves an odd number of sites and since the leads are at half-filling \cite{hofstetter01,*zitko06,*zitko06b,*lopez07,*zitko07,*zitko12}.
When $\Delta V$ is applied, the new rungs in the equivalent system only marginally affect the correlations.

Note that in Reference~[\onlinecite{busser13a}] we were interested in what comes off during the steady state, and we neglected to study the transient regime. Now we want to focus on a different effect that occurs during the transient regime in a case when one of the dots is decoupled before applying the bias. We discuss a such case in the next sections.
 
\begin{figure}
\centerline{\epsfxsize=8.5cm\epsfbox{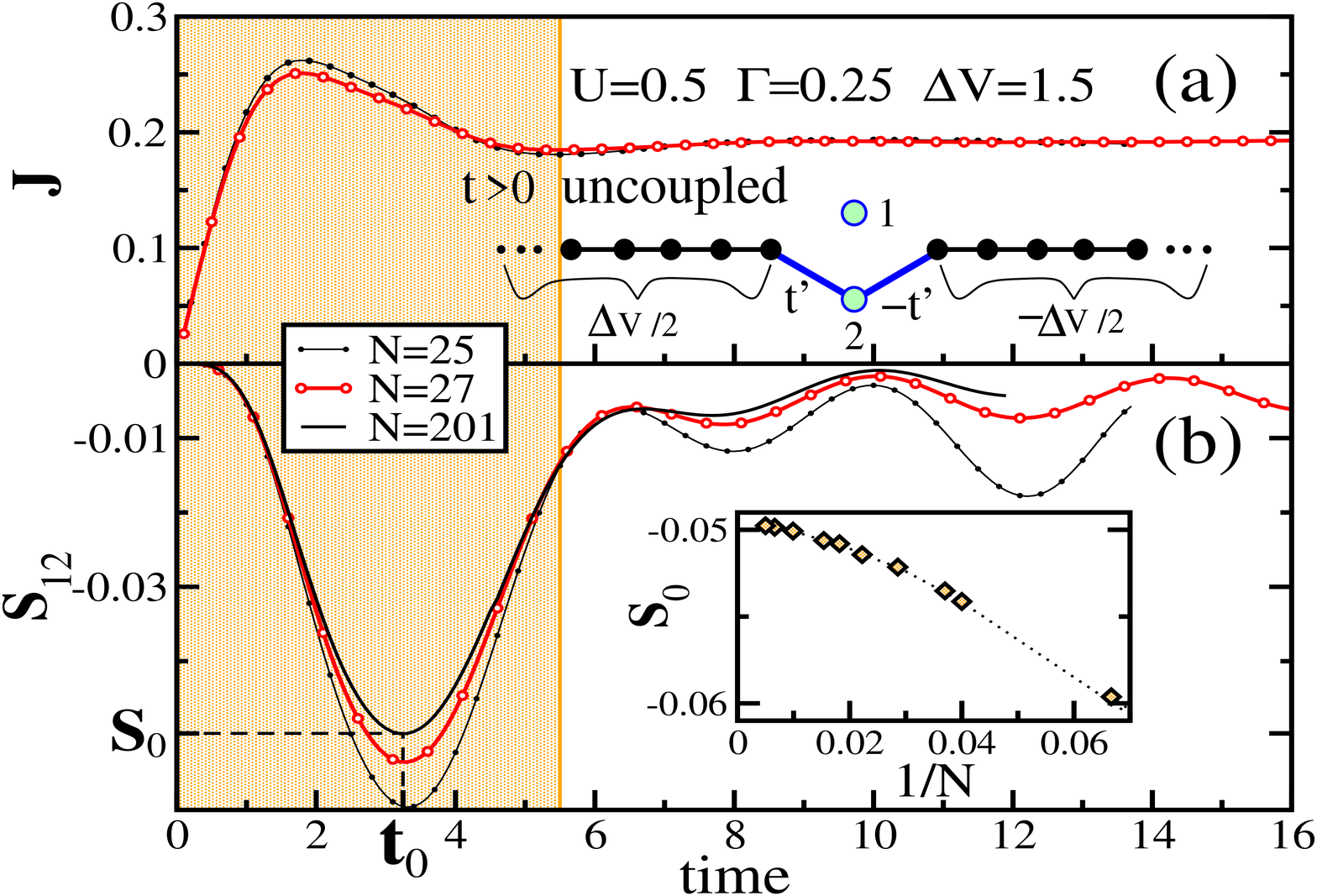}}
\caption{(Color online) Results for current $J(t)$ and spin correlations $S_{12}(t)$ as a function of time for the system with one of the dots uncoupled. In panel (a) we present current as a function of time; as before, a steady current is found after a transient. Observe the shorter transient time due to the absence of the Aharanov-Bohm interference. Panel (b) shows the spin correlation between the 2QDs. The shaded region indicates the transient regime for the current (time $<5.5$).  }
\label{figure2}
\end{figure}

\section{Spin correlations for the uncoupled dot}\label{results}

Once the coupled system is understood, we focus on the study of what happens when one of the dots is decoupled before applying the bias potential $\Delta V$. Again, as we are interested in the AF correlations between the dots, we focus just on the case with phase $\varphi=\pi$.
Observe that we use, as the initial state, the same ground state as in the previous section where both dots are connected to different symmetries. 
Now, for this section, the time evolution is done with a different Hamiltonian where the connection to QD1 is removed ($t'_1=0$).
At the time $t=0$ the QDs are connected to different symmetries, as we know from the canonical transformation shown in the previous section. For this reason, at the time $t=0$, they do not interact and the spin correlation between them is zero.
However, at the initial state, a  Kondo singlet appears between each dot and the corresponding reservoir symmetry (Fig.~\ref{figure0}(a.1) ~). Now we are interested in seeing if the information of the orientation of spin of QD1 is preserved in the Kondo cloud within the symmetric reservoir once the QD is uncoupled. We are also interested to know if this information is transferred to the QD2 when the bias potential is turned on. 
In Figure~\ref{figure0}(c.1) we represented the result for the canonical transformation in the new case with $\Delta V> 0$. Observe that the QD1 is not connected and $\Delta V$ appears as rungs connecting the symmetric with the antisymmetric channel. 

Figure~\ref{figure2} presents the typical results for the current and the spin correlation for the uncoupled system. Current $J$ presents a similar behavior as in the previous section. At the time $t=0$ the current is zero and increases for $t>0$ due to the $\Delta V$ applied. After a transient appears a stationary state where $J$ is almost constant. The value of $J$ in the stationary state is slightly larger than the coupled case seen in the previous section. Also, the transient time is shorter, roughly from time~$\sim 8$ to $5.5$. These two results are due to the absence of the Aharanov-Bohm effect that was present in the previous case. As there are no two paths that interfere, the current is larger, and also the steady state is reached before.

 Even if the results for the current are similar, we found that $\mbox{S}_{12}$ have a different behavior than the results shown in the previous section.
 As before, $\mbox{S}_{12}=0$ at time zero as it is an asset of the initial state as discussed in the previous section. Then, for time $t>0$, the $\mbox{S}_{12}$ develop an AF correlation. This time the spin correlation does not achieve a finite stationary state as before but decays, going to zero for a large time, in the stationary regime of the current.
 This result can be understood considering that the current destroys the residual Kondo cloud formed at the initial state between QD1 and the symmetric lead. It is interesting to note that the residual Kondo cloud can still generate an AF correlation for a short time. 
 
 \begin{figure}
\centerline{\epsfxsize=7.5cm\epsfbox{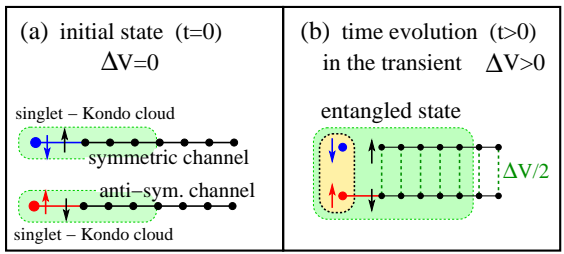}}
\caption{(Color online) Canonical transformation for the uncoupled system. In panel (a) we can see the result of the transformation, Eqs.~\ref{eq:trafo1} and~\ref{eq:trafo2}. At the time $t=0$ both QDs are connected to different reservoirs with different symmetries. Panel (b) show the case for time $t>0$ when the first QD is uncoupled from its reservoir and the bias $\Delta V$ connects both symmetries. In this case, as the even symmetry reservoir is entangled with QD1, and now both reservoirs are connected, a spin correlation between the two dots is generated.  }
\label{figure_explain}
\end{figure}

Even if seems strange to have an AF $\mbox{S}_{12}$ when one of the QDs is uncoupled, this effect can be understood using the canonical transformation and the fact that there was a Kondo effect at the time $t=0$.
In Figure~\ref{figure_explain} we represent schematically the effect of the transformation at $t=0$ and $t>0$.
For $t=0$, the initial state, we have both dots connected to their corresponding symmetric or antisymmetric channels. This connection between a localized spin and a fermionic bath produces the well-known Kondo effect and its Kondo cloud inside the reservoir.\cite{busser10,ribeiro14,ghosh14,feiguin17,laercio19,borzenets20}
For $t>0$, when the quantum QD1 is uncoupled, and the bias is applied ($\Delta V>0$), some of the electrons in the symmetric channel still keep the  Kondo cloud information. The bias potential connects (as rungs) the reservoirs with different symmetries. Even though the QD1 is not connected, the information of its spin projection is transferred to QD2 by the applied bias $\Delta V$.

Another important effect occurs at this point. As a current is circulating from the left to the right lead, a decoherence process starts. As a consequence of this, the entanglement between QD1 and the symmetric reservoir is lost ({\it i.e.} the Kondo cloud is destroyed) and the correlation $\mbox{S}_{12}$  decays for a large time as shown in Fig.~\ref{figure2}(b).
This effect is studied with more detail in reference~\onlinecite{busser14} using tDMRG and also master equation techniques.
 
 We characterize this effect by two values, the larger AF correlation found, $S_0$, and the time when it is achieved $t_0$; both are indicated in the figure. As we mentioned before, in the transient, there is a strong size dependence. This size effect is because the system is not big enough to contain the Kondo cloud. Then the correlation $S_0$ varies strongly with the size of the system.
In the inset of  Fig.~\ref{figure2}(b) we show the detail of the convergence of $S_0$ versus the inverse of the system size $N$. We use for the rest of this article a large value of $N=201$ to get a converged value of $S_0$. 
Observe in the inset of Fig.~\ref{figure2}(b) that $S_0$ converge to the value $\sim -0.05$, which is approximately one-fifth of the value obtained for the connected system in the stationary regime, then the entanglement is smaller than in the previous case.

\begin{figure}
\centerline{\epsfxsize=8.0cm\epsfbox{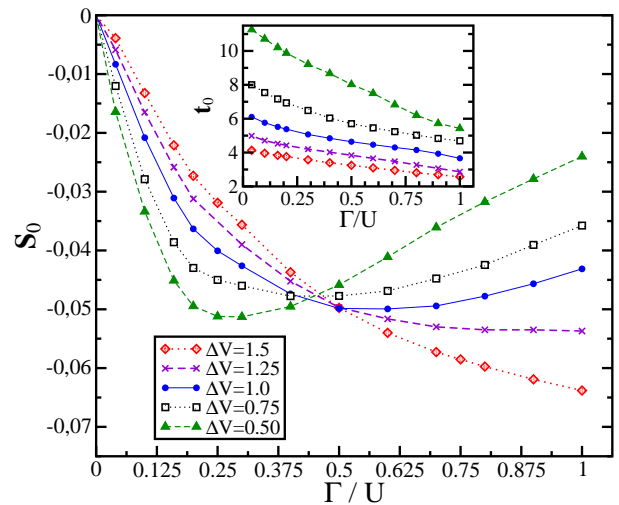}}
\caption{(Color online) Results for the minimum spin correlation $S_0$ found in the transient as a function of $\Gamma/U$ for a fixed $\Delta V$. Note that all the curves intersect close to $\Gamma/U\sim 0.5$; this point separates the cotunneling from the mixed valence regimes. The inset (b) shows the time $t_0$ where $S_0$ is obtained. }
\label{figure5}
\end{figure}

To finish this section, we present in Figure~\ref{figure5} the results of $S_0$ versus $\Gamma/U$ for different $\Delta V$. We can observe that a typical curve has an $S_0$ that goes to zero when coupling factor $\Gamma \to 0$.
This is not surprising, as if the coupling goes to zero (is no interaction between QDs and the reservoirs) cannot appear a spin correlation between the spins located at the QDs.
Next, we observe that the curve has a minimum value and then, for large values of $\Gamma/U$, it decreases again.
The curves for the different values of potential $\Delta V$ intersect at almost the same point close to $\Gamma/U\sim 0.5$. 
In the inset of the figure, we present the value of $t_0$, the time when $S_0$ was achieved. We can observe that it decreases when $\Gamma/U$ increase.
These results can be understood thinking that the current through the metallic leads {\it ''erase''} the entangled state between QD1 and the symmetric reservoir, as explained before.

The explanation of why all the curves intersect near the same point $\Gamma/U\sim 0.5$ is due to the broadening of the energy levels caused by a finite $\Gamma$.
If $\Gamma/U < 0.5$ levels $Vg$ and $Vg+U$ are well separated. If the bias is not large enough, the current is set by the co-tunneling process of the Kondo effect of QD2. This produces a small current and thus reduces the decay effect in $\mbox{S}_{12}$.
For large values of $\Gamma/U$, the mixed valence regime, there is always a direct tunneling channel producing a larger current circulating through QD2 and thus a faster decay in $\mbox{S}_{12}$.
Note that, for larger $\Delta V$, the connection between the symmetric and the antisymmetric channels is larger. Then the transfer of the information is more efficient in such case.
This effect is also discussed in detail in Ref.~\onlinecite{busser14}.

\section{Stability of the entangled state and the decoherence process }\label{quench} 

In previous sections, we had shown that the Kondo effect can be used to generate an entangled state between the spins of two QDs just by applying a bias $\Delta V$ between two metallic leads.
Even more important, we have shown that if the connection to one of the Qds had been removed before the application of $\Delta V$ the entanglement can be generated with the spin information that remains in the contact.
In section~\ref{results} we prepared the initial state to have a Kondo effect at each QD with their respective lead. Then we removed the connection between QD1 and the metallic leads at the same moment the bias was applied.
It is obvious that the process of vanishing the Kondo cloud of the Kondo cloud begins as a consequence of the removal of $t_1$. Then it is interesting to see what happens when we let a certain amount of time elapse between the disconnection of the QD1 and the application of the bias potential. This can be done by changing the Hamiltonian during the time evolution of the DMRG calculations.

\begin{figure}
\centerline{\epsfxsize=8.5cm\epsfbox{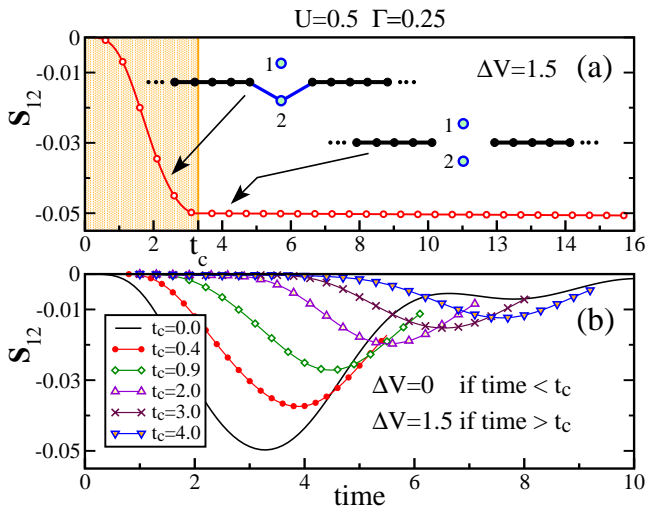}}
\caption{(Color online) The behavior of the spin correlation $S_{12}$ under quantum quenches. In panel (a) we present the stability of the entangled state. We remove the $t_2$ connection at $t_c\sim3.3$ when correlation $S_0$ is achieved. The entangled spins of the QDs do not change as they form an isolated system.
In panel (b) we present the transient of the quench where $\Delta V$ is applied after removing the connection $t_1$ at $t=0$. As soon as $t_1$ is removed, the Kondo cloud starts to vanish. At the time $t=t_c$, when the bias is applied, part of the information of the spin projection of QD1 is lost. As a consequence, the $S_0$ achieved is smaller.  }
\label{quenchtime}
\end{figure}

Before presenting this quenching, we want to show what happens if the connection to the QD2 is also removed when the correlation $S_0$ is achieved. This means changing the time evolution Hamiltonian at the time $t_0$ to have hopping elements $t_2=t_1=0$.
In Figure~\ref{quenchtime}(a) we present this case for the same parameter values of Fig.~\ref{figure2}.
We can see that the small entanglement achieved $S_0$ is maintained due, as is evident, to the two QDs are now isolated. Then the spin correlation between the dots is frozen as there is no mechanism to break the entanglement. This shows that even if the entanglement is small, it is stable and the relative information about the spin projection is preserved.

\begin{figure}
\centerline{\epsfxsize=8.0cm\epsfbox{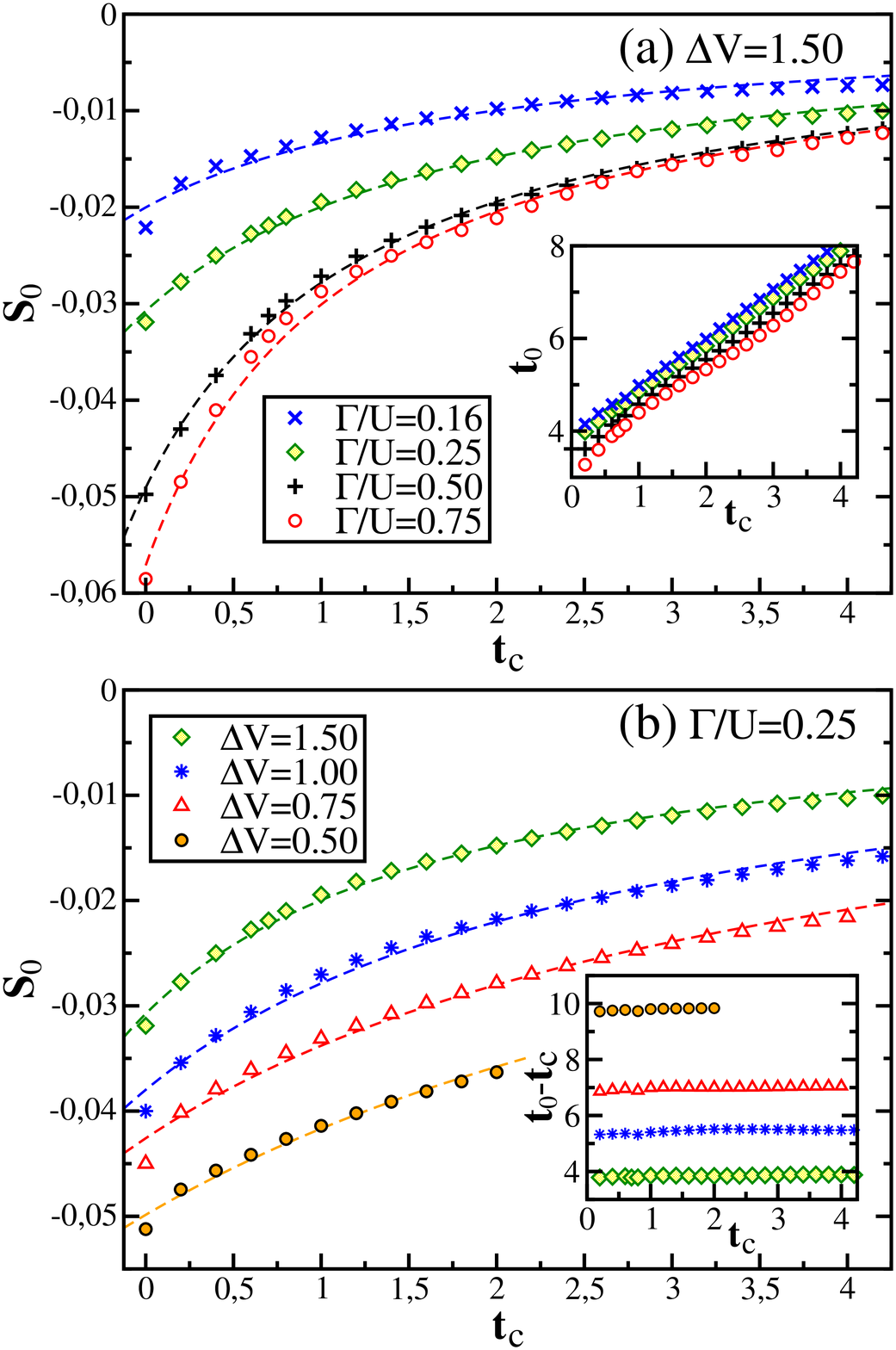}}
\caption{(Color online) Results for the minimum spin correlation $S_0$ for different quenching times $t_c$. In panel (a) we have fixed the bias potential $\Delta V=1.5$ and have calculated $S_0$ for several values of $\Gamma/U$ while in panel (b) $\Gamma/U$ is fixed to $0.25$ and have taken different values for $\Delta V$. The dashed lines are the interpolated curves with the form $\alpha/(t_c-\beta)$ showing a regular decay for $S_0$ as a function of the quenching time $t_c$. The insets present $t_0$ vs. $t_c$ for both cases. }
\label{figure7}
\end{figure}

As we mentioned at the beginning of this section, as soon as connection $t_1$ is suppressed,  the information about the orientation of the spin of the quantum dot 1 begins to vanish. The wiping out of this information is due to the interaction of the Kondo cloud with the rest of the electronic sea. To understand this decoherence process, we will proceed as follows.
First at $t=0$, after calculating the ground state with both QDs connected, we remove the hopping element $t_1$ uncoupling QD1 from the rest of the system. With this new Hamiltonian, we do the temporal evolution of the system, enabling the decoherence process.
Then, at time $t_c$, we apply the bias potential $\Delta V$ connecting the symmetric and the antisymmetric leads. In this way, the QD2 can access the remaining spin information of QD1 allocated at the symmetric reservoir.
As in Section~\ref{results}, during the transient, a small AF correlation appears between the two QDs. In Figure~\ref{quenchtime}(b) we present the results for $S_{12}$ for different values of the switching time of the bias potential $t_c$. For larger values of $t_c$, we obtain smaller values of $S_ {12} $ consistent with the decoherence process of the symmetric channel.
 As there is less information on the spin orientation of QD1 in the symmetric lead, the spin correlation $S_ {12} $ is smaller overall, resulting in a small $S_0$ as can be seen in the figure.

As in previous sections, to characterize this process we use $S_0$, the largest AF correlation achieved during the transient. In Figure~\ref{figure7} we present $S_0$ versus $t_c$ in two situations. 
Panel (a) presents the case where is fixed $\Delta V=1.5$ and took different values of $\Gamma$ while panel (b) we had fixed $\Gamma/U=0.25$, and we took different values of $\Delta V$. In all cases, we interpolate a curve of the form $f(t_c)=\alpha/(t_c-\beta)$ to highlight the monotonic decay.  
The insets inside each panel show the relationship between $t_0$, the time when $S_0$ is achieved, vs. $t_c$. In panel (a) we can appreciate the linear relationship between $t_0$ and $t_c$, while in panel (b) we show that $t_0$ is simply dislocated in $t_c$. 
In all cases, $S_0$ decays monotonically with increasing $t_c$. This is a clear sign that the interaction between the Kondo cloud and the rest of the electron sea dilutes the memory of the QD1 spin projection. This erasure process reduces the maximum correlation $S_0$ reached.
This result is consistent with that found in previous work.\cite{busser14} 

\section{Conclusions}\label{conclusion} 

We demonstrated that spin correlations between separated electrons localized in two different quantum dots can be induced and modified by a bias potential applied between two metal leads connected to the dots in an Aharonov-Bohm interferometer structure.
This spin correlation represents a partially entangled state between the dots.

A small entangled state can still be reached even if one of the QDs is disconnected from the rest of the system before the application of the bias potential. In this case, the entanglement is acquired through the remaining {\it Kondo cloud} formed at the initial state to screen the (now) uncoupled quantum dot. The information about the spin orientation of the uncoupled dot then is transmitted from the electron reservoir to the one that remains coupled.
In this way, these results show that it is possible to use the Kondo entanglement to transfer the spin information from one quantum dot to another to create a new entangled state.

This effect can only occur during the transient. As soon as the bias is applied, a circulating current through the metal contacts starts to destroy the structure of the Kondo cloud and the information about the spin projection starts to be erased. 
The current forms a mechanism that erases the entanglement between the uncoupled quantum dots.

Finally, we have shown that if, during the transient, the connection to the second dot is also removed the partially entangled state achieved is fixed as both dots are isolated ({i.e.} the spin correlation is frozen).

{\bf \it Acknowledgments - } We thank M. Zarea for helpful discussions and point out the direction of this work. 
This work has been supported by the  {\it SeCyT-Rectorado of the University of Mor\'on} under grants { PIO 2020 - 800 201903 00033 UM and PIO 2021 - 800 202101 00033 UM}.

\bibliographystyle{apsrev4-1}
\bibliography{bibliography}

\end{document}